\documentclass[11pt]{article}
\begin{document}
\begin{titlepage}
\begin{center}
{\bf Comment on `Stable Quantum Computation of Unstable Classical Chaos'}
\end{center}
\vskip.2cm
In a recent Letter [1], Georgeot and Shepelyansky have discussed a certain 
coherent quantum simulation of the classical Arnold map. They claim that 
this `classical chaotic system can be simulated on a quantum computer with 
exponential efficiency compared to classical algorithms'. The Comment will 
question their statement. I argue that, as long as the classical evolution 
is concerned, the classical algorithm can be made exactly equivalent
with the quantum one.  

Following the Letter, consider the discretized classical Arnold map.
Start a classical trajectory from $i,j$ and let $i',j'$
denote the new coordinates after many iterations including the 
time-reversal on half-way. The full protocol of the Letter's quantum 
simulation goes obviously like this. First, the quantum amplitudes 
$a_{ij}$ are introduced to represent the initial classical phase-space 
density $\vert a_{ij}\vert^2$. Second, the quantum algorithm is perfomed 
on the amplitudes, using a polinomial number of simple quantum gates. In
the ideal case of perfect gates the resulting unitary transformation reads:
\begin{equation}
a_{ij}\longrightarrow a_{i\prime j\prime}~~.
\end{equation}
Third, quantum measurements are performed regarding those quantum 
observables which do possess interpretation for the classical Arnold 
system as well.

The point is the latter restriction. All information on the final state 
of the classical Arnold system has been encoded into the squared moduli 
$\vert a_{i\prime j\prime}\vert^2$. The complete classical information can
thus be obtained by measuring the projectors
\begin{equation}
\vert x_k\rangle\vert y_l\rangle\langle x_k\vert\langle y_l\vert 
\end{equation}
simultaneously for each $k,l=1,2,\dots,N$. I emphasize that no 
interpretation exists within the framework of the classical Arnold system 
for observables which are not diagonal in the basis 
$\{\vert x_k\rangle\vert y_l\rangle;k,l=1,2,\dots,N\}$.
Any simulation of nondiagonal observables 
would be redundant for the classical evolution [2]. Now, I am going to
show that the measured statistics of the classically relevant quantum 
observables (2) can equivalently be simulated by a classical algorithm 
whose logic steps are just identical with that of the quantum algorithm.

I propose a trivial classical protocol. First, we generate a random pair 
$i,j$ with probability represented by the given initial phase-space density. 
Second, the classical algorithm is performed on $i,j$, using the sequence of 
simple classical gates each being the classical equivalent of the 
quantum gates in the Letter's quantum algorithm. With perfect gates
this leads to $i',j'$. Third, we read out (trivially) the contribution of 
the result to the phase-space density in the final state. Obviously, this 
contribution will be of the same statistics which we would have obtained in 
quantum quantum measurements of observables (2) after quantum computation 
(1). No one could distinguish between the data taken from the quantum or 
from the classical computers, respectively. 

This equivalence remains valid if the coarse-grained or the 
Fourier-trans\-formed version of set (2) is analysed. Furthermore, the 
equivalence survives if logical gates are not perfect. Assume you have a 
classical computer to performe the map $i,j\rightarrow i',j'$, using simple 
reversible gates. And imagine that, at your alternative wish, you can run 
the same gates coherently. This is how the Letter's quantum algorithm  
can be related to the classical one. I have already proved that the 
coherent and incoherent runs give the same statistics for the classical 
Arnold map, provided the gates are perfect. If they are not, we can still 
assume that the bit-error rates are independent of whether we run the 
gates coherently or not [4]. Hence the quantum and classical computations 
will be equivalent for nonideal gates, too. (The gates' phase errors do 
not influence the results of the quantum protocol.) 

For the classical chaotic evolution, the claimed advantage of the 
Letter's quantum algorithm is illusory. It has disappeared when we have 
concretized the statistical analysis, left undetailed by the authors, 
of the final quantum state.

My work was supported by the Hungarian OTKA Grant 032640.
\vskip.3cm
\leftline{~~Lajos Di\'osi}
\leftline{Research Institute for Particle and Nuclear Physics}
\leftline{H-1525 Budapest 114, POB 49, Hungary}

\vskip.3cm
\parindent=0pt

[1] B. Georgeot and D.L. Shepelyansky Phys. Rev. Lett. {\bf 86}, 5393 (2001).
\hfill

[2] The quantum Fourier transform, barely mentioned in [1], is non-diagonal. 
It yields the classically irrelevant power spectrum of the wave function
$a_{ij}$, not the classically relevant harmonics of the Liouville density 
$\vert a_{ij}\vert^2$. For detailed criticism see Ref.~[3].\hfill

[3] A.M. van den Brink, quant-ph/0112006.\hfill

[4] The Letter assumes different errors for quantum and classical gates,
for detailed criticism see Ref.~[5].\hfill

[5] C. Zalka, quant-ph/0110019.\hfill

\end{titlepage}
\end{document}